\documentclass[final]{aipproc}  

\usepackage{graphicx}
\usepackage{bm}
\usepackage{amsmath}
\usepackage{amsfonts}
\usepackage{amssymb}
\usepackage{psfig}

\def\L{\Lambda}

\def\R{\right} \def\L{\left} \def\Sp{\quad} \def\Sp2{\qquad}

\layoutstyle{6x9}

\begin{document}

\title{The nucleon and the nuclear force in the context of effective
  theory and path-integral methods}

\author{L.J. Abu-Raddad\thanks{Email: laith@rcnp.osaka-u.ac.jp}}{address={Theory Group, Research
Center for Nuclear Physics, Osaka University, 10-1 Mihogaoka, Ibaraki
City, Osaka 567-0047, Japan},email={laith@rcnp.osaka-u.ac.jp}}

\begin{abstract}
The nucleon structure and the nuclear force are
investigated in the context of the non-perturbative path-integral
method of hadronization. Starting from a microscopic quark-diquark model, the nucleon is generated as a
relativistic bound state and an effective chiral
meson-nucleon Lagrangian is derived. Many of the nucleon
physical properties are studied using a theory of at most two
free parameters.
\end{abstract}

\maketitle


\section{Introduction}

The central challenge in nuclear physics remains to understand the
origin and nature of the nuclear force due to our inability to solve
quantum chromodynamics (QCD), the fundamental theory for the strong
interactions.  The basic problem of QCD is that its fundamental degrees
of freedom, quarks and gluons, are not the observable baryon and meson
states. Thus bridging the missing link between the fundamental and
observable degrees of freedom stands as one of the stark challenges of
nuclear/elementary particle physics today. Although we do have an ab
initio approach to solve this problem, that is lattice QCD, this
endeavor is still miles away from achieving such a goal. For the time
being, we have no alternative but to resort to effective non-perturbative
approaches of which this study is one.

This presentation describes our work~\cite{AHTE02} in addressing this missing link by
deriving a chiral meson-nucleon Lagrangian from a microscopic model of
quarks and diquarks using the path-integral method of hadronization. Chiral symmetry and
its spontaneous breaking have proven to be key concepts
in understanding meson and baryon structure and many features of the
nuclear force.  Therefore, we start from a QCD-based chiral effective field
theory, the Nambu Jona-Lasinio (NJL)
model that accommodates most of QCD symmetries~\cite{HK94}. 
Then, the nucleon is described as quark-diquark correlations. This
assertion is vindicated by a mounting experimental evidence that
diquarks play a dynamical role in hadrons~\cite{APEFL93}.
Using the path integral hadronization, we calculate nucleon properties
and derive an effective chiral meson-nucleon Lagrangian of the quantum
hadrodynamics (QHD) type~\cite{SW86} that describes the rich
meson-nucleon interactions in a fully covariant and chirally symmetric
formalism.

While this program is applied to the case of nucleons and mesons, it
is certainly of general nature and can possibly be applied
prolifically to yield other baryons and their interactions.  Moreover,
the idea of using path-integral techniques to transform a Lagrangian
from its fundamental to its composite degrees of freedom is a powerful
concept in physics of immense impact and utility. As a matter of fact,
the authors of Ref.~\cite{IMO01} have recently invoked such techniques
in their study of high-temperature superconductivity.  They succeeded
in doing so by converting a model of strongly-correlated electrons
into an effective U(1) gauge field theory in terms of composite
fields.

The use of hadronization has been introduced in Ref.~\cite{Cahill89, Rein90}. Consequently, the authors of Ref.~\cite{EJ98} attempted to construct an effective Lagrangian for the nucleon using only scalar
diquarks.  In the present work, we extend their work by deriving the
structure using both axial-vector and
scalar diquarks and we employ, as opposed to Ref.~\cite{EJ98}, a
gauge-invariant regularization scheme throughout our analysis.
Furthermore, we verify the Ward identity and the Goldberger-Treiman
relation and present a full numerical study of various nucleon
observables for the case of scalar diquarks drawing special attention
to the role of an intrinsic diquark form factor. 
Hence, this work is the first calculation of an extensive set of nucleon
observables using path-integral hadronization since the introduction
of the idea more than ten years ago.

\section{Formalism}

We start from an NJL Lagrangian satisfying SU(2)$_L\; \times$
SU(2)$_R$ chiral symmetry:
\begin{equation}
{\mathcal L}_{NJL} = {\bar q} (i\rlap/\partial -m_0) q + \frac{G}{2} \left[
({\bar q} q)^2 + ({\bar q} i \gamma_5 \vec{\tau} q)^2 \right]\;,
\end{equation}
where $q$ is the current quark field, $\vec{\tau}$ are the Pauli matrices, $G$
is the NJL coupling constant, and $m_0$ is the current quark mass
which explicitly breaks chiral symmetry. The color and flavor indices
are suppressed for brevity.

By introducing composite scalar ($\sigma \sim \bar q q$) and
pseudoscalar ($\vec{\pi} \sim \bar q i \gamma_5 \vec \tau q$) fields through the 
Hubbard-Stratonovich transformation~\cite{ERV94},  
we can rewrite the NJL lagrangian into a semi-bosonized Lagrangian
${\mathcal L}^\prime_{NJL} = \L[\bar q (i\rlap/\partial -\sigma-i\gamma_5
\vec\tau\cdot\vec\pi)q -\frac{1}{2G}(\sigma^2+\vec\pi^2)\R]$, where we have absorbed the bare quark mass 
$m_0$ into the sigma field $\sigma$.  
Further, we transform the meson fields
according to the non-linear parameterization $\L[\sigma,\pi\R]
\rightarrow \L[\sigma^\prime,\Phi\R]$:
$\sigma+i\gamma_5\vec\tau\cdot\vec\pi=\left( m_q +
\sigma^\prime\right) {\rm e}^{-\frac{\rm
i}{F_\pi}\gamma_5\vec\tau\cdot\vec\Phi}$.
Here, $F_\pi = 93$~MeV is the pion decay constant and $m_q\equiv
\langle\sigma\rangle_0$ is the constituent quark mass which is fixed
through a gap equation in the meson sector~\cite{HK94}.

As a consequence of considering Lorentz structure, there are five
types of possible diquark $qq$ correlations. These are the scalar, pseudo-scalar, vector, axial-vector, and
tensor diquarks. For each of these Lorentz
$qq$ formations, we have also an isoscalar and isovector 
diquarks. Using permutation
symmetry and Fierz transformation, we have verified an earlier assertion~\cite{EPT83}
that only two diquark formations are independent for the nucleon if
its field is
to be written as a local operator of three quarks. 
Hence, we introduce ${\vec D}^{\mu}$ as
an axial-vector isovector diquark field and $D$ as a scalar isoscalar one.

Next we introduce a quark-diquark interaction term in such a manner to
generate the nucleon as a linear combination of axial-vector and scalar
diquarks. It is convenient here, considering chiral symmetry, to work with the chirally rotated $\chi$
``constituent'' quark field defined by $\chi \equiv
e^{\frac{-i}{F_\pi} \gamma^5 \frac{\vec{\tau}}{2} \cdot \vec{\Phi}}
q$. By introducing electromagnetic interactions and batching the
semi-bosonized NJL Lagrangian, the diquark contributions and the
quark-diquark interaction term, 
we obtain the following Lagrangian as
our microscopic model:
\begin{eqnarray}
{\mathcal L} &=& \bar{\chi}\; S^{-1} \; \chi 
\;-\; \frac{1}{2G}(\sigma^{\prime}+m_q)^2
\;+\; \delta{\mathcal L}_{\rm sb}
\;+\; D^\dag \;\Delta^{-1} \;D \;+\; {\vec{D}^{\dag\;\mu}} \;
{\tilde{\Delta}}^{-1}_{\mu \nu} \; \vec{D}^{\nu} \;+\; \nonumber\\ &&
\tilde{G} \L( \sin{\theta} \; \bar{\chi}\gamma^\mu \gamma^5
\:\vec{\tau}\cdot {\vec D}^\dag_{\mu} \;+\; \cos{\theta}
\;\bar{\chi} D^\dag\R)\; \L(\sin{\theta}\; {\vec D}_{\nu}\cdot
\vec{\tau} \: \gamma^\nu \gamma^5 \chi \;+\; \cos{\theta} \;D \chi \R)\;,
\label{lsemibos}
\end{eqnarray}
where
\begin{subequations}
\label{definitions1}
\begin{eqnarray}
S^{-1} &=& S_0^{-1} \;+\; {\mathcal M}\;,\\
{\mathcal M} &=& -\left[ \gamma^\mu \frac{\vec{\tau}}{2}\cdot \vec{\mathcal
V}^{\pi}_\mu \;+\; \gamma^\mu \gamma^5 \frac{\vec{\tau}}{2}\cdot
\vec{\mathcal A}^\pi_\mu \;+\; \sigma^\prime \;+\; \gamma^\mu Q_q
A^{\rm EM}_{\mu} \right]\;,\\
\Delta^{-1} &=& \Delta_{0}^{-1} + i Q_S A^{\rm EM}_{\mu}
(\overrightarrow{\partial^\mu} - \overleftarrow{\partial^\mu})\;,\\
{\tilde{\Delta}}^{-1}_{\mu \nu} &=& {{\tilde{\Delta}}_0\;}^{-1}_{\mu
\nu} \;+\; i Q_A \left[ (A^{\rm EM}_{\mu} \overleftarrow{\partial_\nu}
- A^{\rm EM}_{\nu} \overrightarrow{\partial_\mu} ) \;-\; g_{\mu \nu}
A^{\rm EM}_{\alpha}(\overleftarrow{\partial^\alpha} -
\overrightarrow{\partial^\alpha}) \right]\;.
\end{eqnarray}
\end{subequations}
Here $\theta$ is a mixing angle for the two diquark contributions,
$\tilde{G}$ is the quark-diquark coupling constant and $Q_q$, $Q_S$
and $Q_A$ are the quark and diquark charges while $S_0$, $\Delta_{0}$
and ${{\tilde{\Delta}}_0\;}_{\mu \nu}$ are the free quark and diquark
propagators (Notice that ${\mathcal O}(Q_{S,A}^2)$ terms are discarded in
Eq.~(\ref{definitions1})). The ${\mathcal M}$ matrix contains all interaction
vertices of the quark field with meson and electroweak fields (weak
part is not shown), where the vector $\vec{\mathcal V}^{\pi}_\mu$
and the axial vector $\vec{\mathcal A}^\pi_\mu $ fields are defined
through the Cartan decomposition ($\vec{\xi} \equiv
\frac{\vec\Phi}{F_\pi}$): ${\exp}\L({-\frac{\rm
i}2\gamma^5{\vec\tau\cdot\vec\xi}}\R) \;\partial_\mu\
{\exp}\L({\frac{i}{2} \gamma^5{\vec\tau\cdot\vec\xi}}\R) \;=\;
{\textstyle\frac{i}{2}\; \gamma^5\;{\vec\tau} }\cdot\vec {\mathcal
A}^\pi_\mu(\xi)\;+\; {\textstyle\frac{i}{2} \;
{\vec\tau}}\cdot\vec{\mathcal V}^\pi_\mu(\xi)$.

Subsequently, we introduce collective nucleon fields ($B \sim \sin{\theta} {\vec
D}_{\nu}\cdot \vec{\tau}  \gamma^\nu \gamma^5 \chi +
\cos{\theta} D \chi$) through another
Hubbard-Stratonovich transformation.
At this point only quarks and
diquarks are dynamical fields with kinetic terms while the meson and nucleon fields are merely
auxiliary ones. By integrating over the quark and then over
the diquark fields, we obtain a meson-nucleon effective Lagrangian. 
Accordingly, we have
``hadronized'' the microscopic theory by producing the dynamical meson
(bosonization) and nucleon (fermionization) fields.

Thereupon, we arrive at a compact Lagrangian given by
\begin{eqnarray}
{\mathcal L}_{\rm eff} &=& \delta{\mathcal L}_{\rm sb} 
\;-\; \frac{1}{2G}(\sigma^{\prime}+m_q)^2
\;-\; i \;{\rm tr\; ln}
S^{-1} \;-\; \frac{1}{\tilde{G}}\; \bar{B} B \;+\; i\;{\rm tr \; ln} (
1 \;-\; \Box ) \;+\; \nonumber \\ &&i\; {\rm \;tr\; ln} ( 1 \;-\;
\Delta_0\; {\rm EM}\;{\rm Int}) \;+\; i\;{\rm tr\; ln} ( 1 \;-\;
\tilde{\Delta}_0 \;{\rm EM}\;{\rm Int})\;.
\label{effL}
\end{eqnarray}
Here the trace is over color, flavor and Lorentz indices while
the ``EM Int'' label stands for the diquark electromagnetic interaction terms. Furthermore,
\begin{subequations}
\label{definitions2}
\begin{eqnarray}
\Box & = & \begin{pmatrix} {\mathcal{A}}& {{\cal{F}}_2}\\
 {{\cal{F}}_1} &{\mathcal S}
\end{pmatrix},\\
{\mathcal A}^{\mu i,\,\nu j} & = & {\sin}^2\theta \bar{B} \; 
\gamma_\rho \gamma^5\;
  {\tau}_{k} \;\tilde{\Delta}^{\rho k,\,\mu i} \;
 S\; {\tau}^{j}\;
  \gamma^\nu \gamma^5\; B\;,\\ 
{\mathcal S}& = & {\cos}^2\theta\; \bar{B}\;\Delta
  \;S\;B\;,\;\; \;\;\;\;\\ 
({{\mathcal F}}_1)^{\nu j}& = & \sin{\theta} \cos{\theta}\;\bar{B}\; 
\Delta \;S\; 
 {\tau^j}\;\gamma^\nu \gamma^5\;  B\;,\\
({{\mathcal F}}_2)^{\mu i}& = & \sin{\theta} \cos{\theta} \;
\;\bar{B} \;\tilde\Delta^{\rho k,\, \mu i}\;\gamma_\rho \gamma^5\; {\tau_k} \; S\; B\;.
\end{eqnarray}
\end{subequations}
This effective Lagrangian contains plenty of rich physics: kinetic and
mass terms for nucleons and mesons together with a multitude of 
interaction terms of mesons, nucleons, and electroweak gauge
bosons. Nonetheless, the
most desired part of the Lagrangian is the prized chiral
meson-nucleon interaction and nucleon-nucleon vertices which delineates the nuclear force.

\begin{figure}
\includegraphics[totalheight=1.0in]{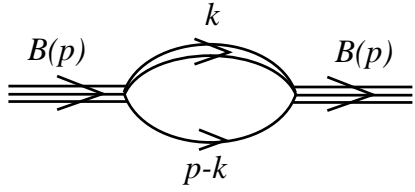}
\caption{\label{fig1} The self-energy diagram which generates the
nucleon kinetic and mass terms and produces the mass equation that
determines the nucleon mass.}
\end{figure}

In order to explore the physics of the nucleon sector, we take the
leading term in the loop and derivative expansion of $i\: {\rm tr\;
\rm ln} ( 1 - \Box ) - \frac{1}{\tilde{G}}\; \bar{B} B$ $\rightarrow$
-$\int {\rm d}^4x  {\rm d}^4y \bar{B}(x)\left[ \Sigma(x,y) \;+\;\frac{1}{\tilde{G}} \delta(x-y)
\right]B(y)$, which is nothing but the nucleon self-energy (see Fig.~\ref{fig1}). The Fourier transform of $\Sigma$ is then decomposed as
$\Sigma(p) = \Sigma_s(p^2) \;+\; \rlap/p \; \Sigma_v(p^2)$.  This
leading term generates dynamically the nucleon
mass $M_B$ which is extracted as the pole of the propagator:
\begin{eqnarray} 
\frac{1}{\tilde{G}} \;+\; \Sigma_s(M_B^2) \;+\; M_B \;\Sigma_v(M_B^2)
= 0\;.
\label{mass}
\end{eqnarray}
Thus, near the mass shell the inverse nucleon propagator takes the
form:
\begin{eqnarray}
\left[ \Sigma(p^2) \;+\; \frac{1}{\tilde{G}}\right] &\sim&
(\rlap/p-M_B) \;Z^{-1}\;,
\end{eqnarray}
where $Z$ is the
wave-function renormalization constant ($B = \sqrt{Z} \; B_{\rm
ren}$). Evidently, the nucleon has finally acquired the desired status as a
dynamical degree of freedom in the problem.

\begin{figure}
\includegraphics[totalheight=1.0in]{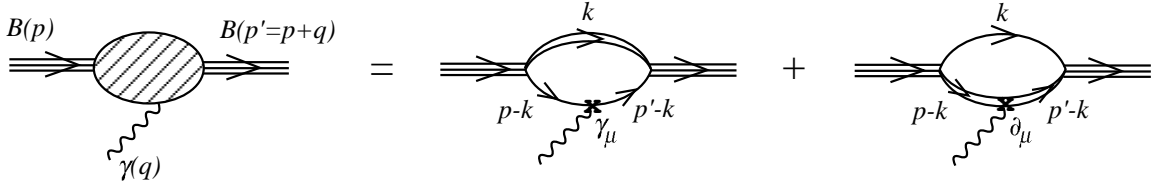}\\
\caption{\label{fig2} The Feynman diagrams for the electromagnetic
coupling which generate the electromagnetic vertex of
the nucleon and subsequently are used to test the
validity of the Ward identity.}
\end{figure}

In computing the various Feynman diagrams in the problem, we encounter
divergent integrals that must be regularized. Several regularization
schemes were attempted. We found that
the most suitable scheme is the PV technique
which we have adopted as the standard method in this work. Accordingly,
we verified the Ward-Takahashi identity by computing the electromagnetic vertex
shown in Fig.~\ref{fig2}.  As a matter of principle, the PV
mass in the nucleon sector can be different from
the NJL cut-off arising in the meson sector~\cite{HK94}. Nonetheless, to minimize the number of free parameters, we
elected to equate them. It is
noteworthy here that all observables were found to be very insensitive to
the value of the PV mass upholding the futility of using it as a free
parameter.

We have also verified one of the chiral symmetry relations, the Goldberger-Treiman
relation, by computing both the axial-vector coupling constant $g_A$
and the pion nucleon coupling constant $g_{\pi N N}$. In the
hadronization formalism, this relation emerges naturally intact as
opposed to large violations in the Bethe-Salpeter equation approach~\cite{HAOR97}

\section{Numerical study}

Having derived the structure of the Lagrangian, we proceed to
generate numerical results using only scalar diquarks ($\theta = 0$ in Eq.~(\ref{lsemibos}) and (\ref{definitions2})), thereby admitting the
possibility of an intrinsic diquark form factor (IDFF).  Including only
scalar diquarks is not out of place as many recent studies using such
diquarks have reported good results for most of the nucleon
observables~\cite{HW95,Keiner96a,HAOR97}. Moreover, there are strong
indications of large scalar diquark dominance in the
nucleon~\cite{MBIY02}.

The parameters $G$ and the cut-off $\Lambda$ are fixed to yield the
constituent quark mass and the pion decay constant~\cite{HK94}. The
diquark masses are also determined in the NJL model~\cite{VW91}. This
leaves us with only one free parameter in our model: the quark-diquark
coupling constant $\tilde{G}$ which is fixed to determine the nucleon
mass of 0.94~GeV through the mass equation~(\ref{mass}). Thus the basic
quantities in our model are the constituent quark mass $m_q = .390$~GeV,
the scalar diquark mass $M_D = .600$~GeV, the quark-diquark coupling constant
$\tilde{G}= 159.1$~GeV$^{-1}$, and the Pauli-Villars mass $\Lambda = .600$~GeV.
We obtain a binding energy of $\Delta E_{\text{bin}} \equiv m_q + M_D - M_B =
50$~MeV, suggesting a loosely bound state for the nucleon.

\begin{table}
\caption{\label{tab2}Some of the nucleon static
properties as predicted in the present calculation using the intrinsic
diquark form factor (IDFF) or without it. Experimental
values are taken from Ref.~\cite{Dumbrajs83,particle00}.}
\begin{tabular}{cccccccc}
\hline
& $\mu_p$ & $\mu_n$ & $g_A$ & $<r^2>^p_E$ & $<r^2>^n_E$
& $<r^2>^p_M$  & $<r^2>^n_M$\\
&&&&(fm$^2$)&(fm$^2$)&(fm$^2$)&(fm$^2$)\\
\hline
Theory with IDFF& 1.57 & -.75 & .87 & .77& -.11& .82& .84\\
Theory without IDFF& 1.57 & -.75 & .87 & .68& -.19& .82& .85\\
Experiment&2.79 & -1.91 &  1.26 & .74  & -.12 & .74 & .77\\
\hline
\end{tabular}
\end{table}

Tab.~\ref{tab2} displays our predictions for some of the static
properties of the nucleon. For the nucleon magnetic moments, our treatment predicts a number that is two-third of the
experimental value for the proton and one-half of that for the
neutron. This is not a surprising result as we have not
included the axial-vector diquark in the present
calculation. 
The predicted value for the
axial-vector coupling $g_A$ of $0.87$ is less than the
experimental one of $1.26$ indicating here also the importance of the
axial-vector diquark.

The nucleon size is nicely well-produced in our model as
the electric and magnetic radii for the nucleon are
very close to the experimental measurements. The negative charge radius of
the neutron has been suggested as an indication of a scalar diquark
clustering in the nucleon~\cite{DMW81} and our treatment dynamically
manifests this assertion. 
These numbers point to a
physical picture of a ``heavy'' diquark at the center with a quark
rotating around it. 
The extended size of the diquark contributes a positive value of about
$0.10~\text{fm}^2$ for the nucleon electric
radii. As expected, the IDFF has virtually no effect on the
magnetic radii as the scalar diquark has a negligible contribution to
the magnetic form factors.

\begin{figure}[t]
\includegraphics[totalheight=5.0in,angle=-90]{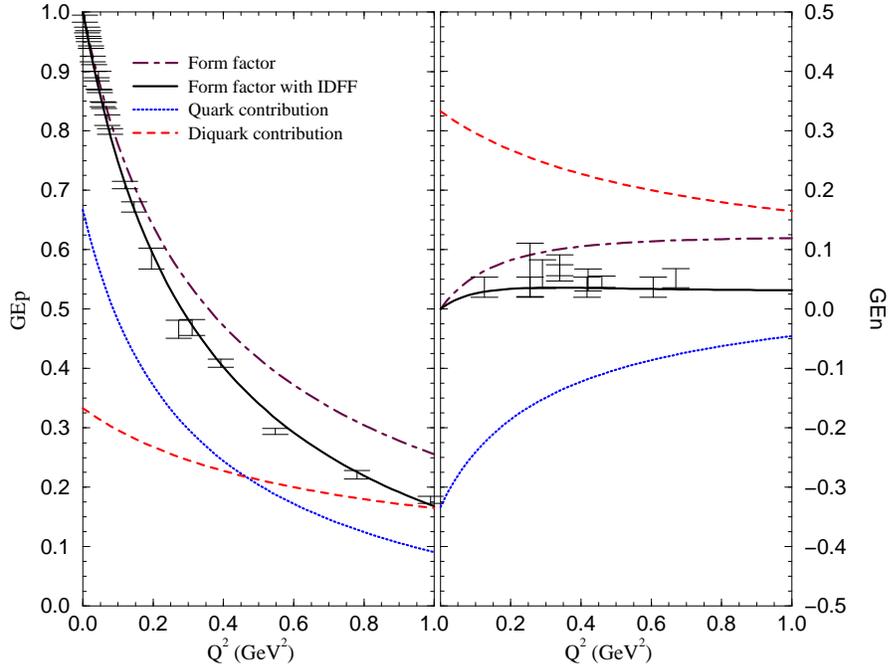}
\caption{\label{fig3} The nucleon electric form factors with and
without the intrinsic diquark form factor (IDFF) along with its quark
and diquark contribution. The left panel shows the proton results
while the right panel displays those for the neutron.  Experimental
data can be found in Ref.~\cite{Hohler76,lomon01}.}
\end{figure}

Next we calculate the nucleon form factors. The left panel of
Fig.~\ref{fig3} displays the proton form factor with and without the
IDFF along with its quark and diquark contribution.
Our treatment produces beautifully this observable. It is evident here
that the IDFF~\cite{WBAR93} plays an important role specially at large
values of momentum transfer ($Q^2 \equiv - q^2$ where $q^\mu$ is the
momentum transfer).  The neutron electric form factor tells a similar
story (right panel).  Clearly, the quark contribution is negative in
value (d-quark) and thus cancels much of the diquark contribution
leading to a small form factor.  It is noteworthy here that the
neutron form factor is a potent test of any treatment as it is a
delicate cancellation of two large
contributions~\cite{Keiner96a}. Saliently, the cancellation is
naturally produced in our study.

\begin{figure}[t]
\includegraphics[totalheight=5.0in,angle=-90]{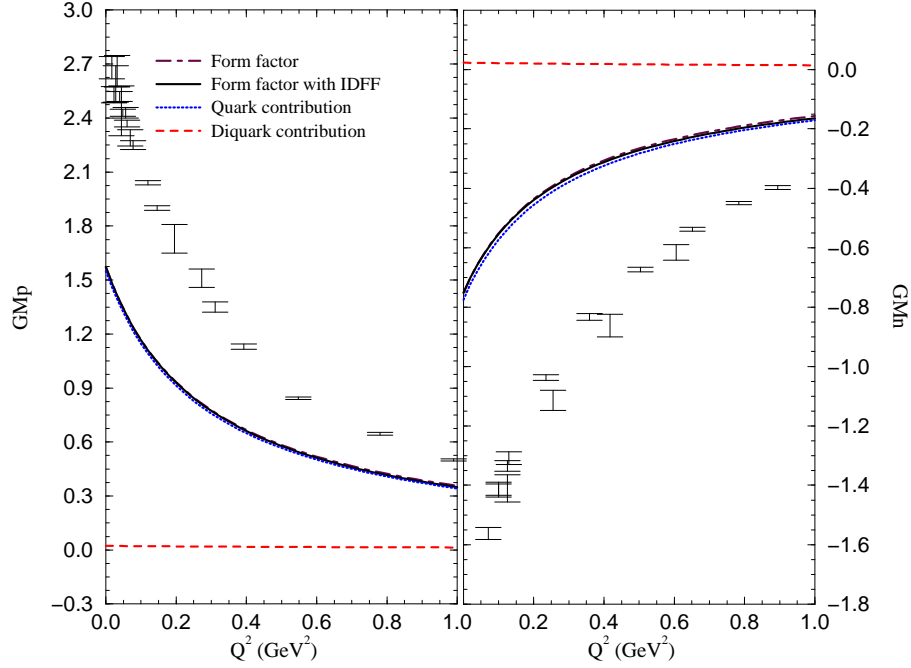}
\caption{\label{fig4} The nucleon magnetic form factors with and
without the intrinsic diquark form factor (IDFF) along with its quark
and diquark contribution. The left panel shows the proton results
while the right panel displays those for the neutron.  Experimental
data can be found in Ref.~\cite{Hohler76,lomon01}.}
\end{figure}

In Fig.~\ref{fig4} we present the nucleon magnetic form factor as
calculated with or without the intrinsic diquark form factor. Unmistakably, the scalar diquark contribution is
virtually vanishing due to the lack of an intrinsic spin.
Nevertheless, there is a very small contribution due to a small
orbital angular-momentum effect in the bound quark-diquark system.
A comparison with experimental data suggests the need for the axial-vector
diquark, which does have an intrinsic spin, to supplement the quark
contributions and to provide the missing strengths for the magnetic
form factors.

\section{Conclusions}

In conclusion, we have tackled the nucleon structure and the
challenging problem of understanding the origin and nature of the
nuclear force by deriving a meson-nucleon Lagrangian using the
path-integral method of hadronization. 
The treatment produced a remarkable agreement with
experimental data for the nucleon size and its electric form factors,
while our calculations show missing strengths for the magnetic properties and $g_A$. The discrepancy is
likely due to the absence of the axial-vector diquark in the present
numerical study.
This presentation describes the first work in our program of using path-integral
hadronization to study baryon structure and
the nuclear force. Deriving this force provides nuclear
physics with a solution to the stigma of no fundamental foundation
that has tarnished its image for decades. 
\begin{theacknowledgments}
I would like to acknowledge the support of a joint fellowship from the Japan
Society for the Promotion of Science and the United States National
Science Foundation as well as the support of all of the sponsors of
the ``New States of Matter in Hadronic Interactions'' conference.
\end{theacknowledgments}

\bibliographystyle{aipproc}   

\bibliography{baryon}

\IfFileExists{\jobname.bbl}{}
 {\typeout{}
  \typeout{******************************************}
  \typeout{** Please run "bibtex \jobname" to optain}
  \typeout{** the bibliography and then re-run LaTeX}
  \typeout{** twice to fix the references!}
  \typeout{******************************************}
  \typeout{}
 }

\end{document}